# High-performance versatile setup for simultaneous Brillouin-Raman micro-spectroscopy

F. Scarponi,[1] S. Mattana,[2] S. Corezzi,[2,*] S. Caponi,[3] L. Comez,[3] P. Sassi,[4] A. Morresi,[4] M. Paolantoni,[4] L. Urbanelli,[5] C. Emiliani,[5] L. Roscini,[6] L. Corte,[6] G. Cardinali,[6] F. Palombo,[7] J. R. Sandercock,[1] and D. Fioretto[2,†]

[1] *Tablestable Ltd., Im Grindel 6, CH-8932 Mettmenstetten, Switzerland*
[2] *Dipartimento di Fisica e Geologia, Università di Perugia, Via Pascoli, I-06123 Perugia, Italy*
[3] *IOM-CNR c/o Dipartimento di Fisica e Geologia, Università di Perugia, Via Pascoli, I-06123 Perugia, Italy*
[4] *Dipartimento di Chimica Biologia e Biotecnologia, Università di Perugia, Via Elce di Sotto 8, I-06123 Perugia, Italy*
[5] *Dipartimento di Chimica Biologia e Biotecnologia, Università di Perugia, Via del Giochetto, I-06123 Perugia, Italy*
[6] *Department of Pharmaceutical Sciences-Microbiology, University of Perugia, Borgo 20 Giugno 74, 06121 Perugia, Italy*
[7] *University of Exeter, School of Physics and Astronomy, Exeter EX4 4QL, UK*

**Brillouin and Raman scattering spectroscopy are established techniques for the nondestructive contactless and label-free readout of mechanical, chemical and structural properties of condensed matter. Brillouin-Raman investigations currently require separate measurements and a site-matching approach to obtain complementary information from a sample.**
**Here we demonstrate a new concept of fully scanning multimodal micro-spectroscopy for simultaneous detection of Brillouin and Raman light scattering in an exceptionally wide spectral range, from fractions of GHz to hundreds of THz. It yields an unprecedented 150 dB contrast, which is especially important for the analysis of opaque or turbid media such as biomedical samples, and a spatial resolution on sub-cellular scale.**
**We report the first applications of this new multimodal method to a range of systems, from a single cell to the fast reaction kinetics of a curing process, and the mechano-chemical mapping of highly scattering biological samples.**

---

[*] Corresponding author. silvia.corezzi@unipg.it
[†] Corresponding author. daniele.fioretto@unipg.it

Brillouin and Raman spectroscopy are inelastic light scattering techniques that, although they only differ in the probed frequency range, give complementary information: mechanical and chemical properties of matter. In spite of their origin around the years 1920s[1,2], the two techniques have been developed as separate tools, also in combination with optical microscopy. Brillouin light scattering (BLS) is the inelastic scattering of light from collective modes, such as acoustic waves (phonons) and spin waves (magnons) propagating in condensed matter, that induce frequency shifts of the radiation in the range 0.1 – 100 GHz. BLS spectroscopy, providing information on the elastic[3], viscoelastic[4] and magnetic properties of matter,[5,6] has been widely applied in condensed matter physics but only sparingly in the biomedical field[7-9], where it can provide a high-frequency counterpart to traditional mechanical techniques[10]. Raman scattering (RS) deals with light inelastically scattered by optical phonons or intra-molecular modes, with frequency shifts typically larger than 1 THz. RS spectra contain a chemical fingerprint of the material with information concerning the molecular composition and structure. RS is widely used in a number of fields ranging from analytical science to biophotonics and biomedical sciences.[11] Recently, the conventional barrier between the two techniques is crumbling and increasing interest is devoted to simultaneous Brillouin and Raman Light Scattering (BRLS) investigations. Extended depolarized light scattering (EDLS) based on separate Brillouin and Raman measurements has been introduced to study solvation processes [12] and complex relaxation patterns in glasses [13] and glass forming materials.[14] Also, Brillouin and Raman microscopy have been applied through a site-matched approach to obtain mechanical mapping with chemical specificity of human tissues[15,16], thus opening the route to a wide range of biomedical and bioengineering applications. Although the contactless nature of light scattering considerably improves the potential for noninvasive real-time, *in-vivo* applications, the full development of Brillouin microscopy has been hampered by long acquisition times (of the order of minutes with traditional Fabry–Perot interferometers).

Advances in Brillouin microscopy have been recently achieved by introducing a non-scanning virtually imaged phase array (VIPA) in place of the Fabry-Perot (FP) interferometer.[9,17,18] The low contrast of VIPA has been improved by multi-pass configurations[19], eventually used in combination with a triple-pass Fabry-Perot as a bandpass filter[20] or using a molecular/atomic gas cell as a notch filter.[21] This solution has also been adopted in a joint BLS-RS setup. [22,23] While VIPAs considerably improve the Brillouin data gathering speed, this comes at a cost: *i*) a rather coarse spectral resolution, which is limited to ~0.7 GHz by the fixed thickness of the etalon; *ii*) a low contrast, which is 30 dB in the single-pass setup and reaches 85 dB in the filter-combined multi-pass configuration; *iii*) a reduced spectral range, which is limited to some tens of GHz by the repetition of orders in the etalon response function. The VIPA-based approach remains limited to moderately turbid media or tissue phantoms, because available spectrometers have not yet achieved the contrast that is required to interrogate truly opaque media such as most biological samples. Furthermore, a better resolution of the spectrometer is needed to accurately measure the linewidth of Brillouin peaks, and to migrate from a purely mechanical characterization (related only to the peak frequency) towards a viscoelastic characterization (related to the peak frequency and linewidth) of biomaterials, including living cells and non-transparent tissues.

In this Article, we present a *high resolution*, *high contrast* and *wide spectral range* confocal micro-spectroscopic set-up based on a new concept of multi-pass FP interferometer and a conventional Raman spectrometer, enabling simultaneous Brillouin-Raman micro-spectroscopy (BRaMS) for a range of applications unapproachable by ordinary devices.

Fig.1(a) shows a schematic of the setup. Visible light from a laser source (S) is focused onto the sample by the same objective that is also used to collect the backscattered light. The sample is mounted onto a three-axes piezo translation stage for mapping measurements (SH). A polarizing beam splitter (BS) transmits the depolarized backscattered light to the spectrometers. Immediately after, a short-pass tunable edge filter (TEF) transmits the quasielastic scattered light to a new concept of tandem Fabry–Perot interferometer (TFP-2 HC, see Methods) and reflects the deeply inelastic scattered light into a Raman monochromator (RM).

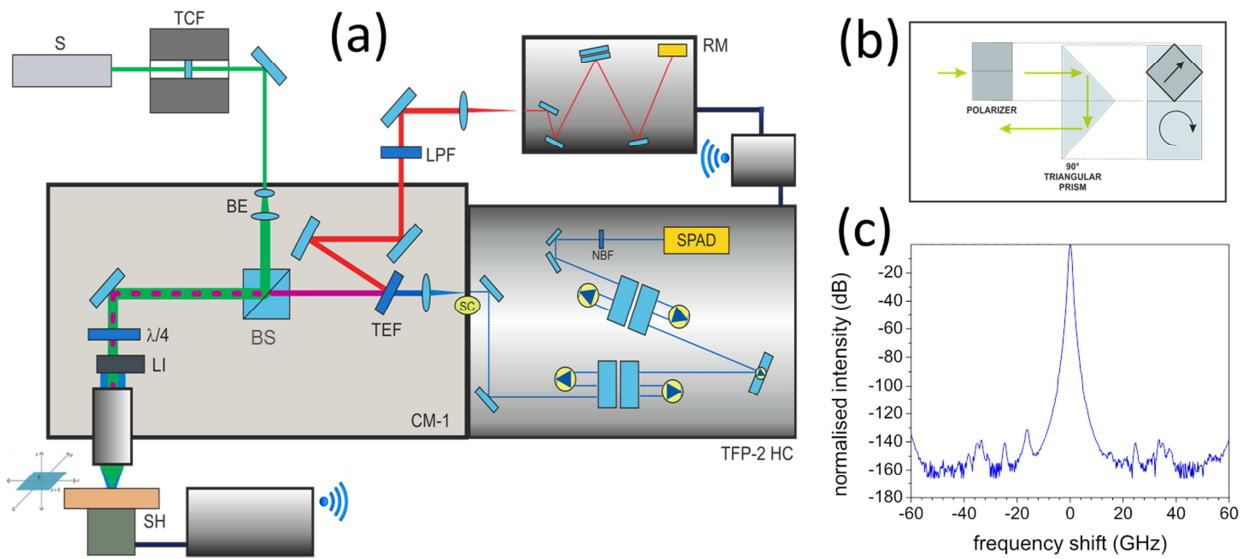

**Figure 1** (a) Schematic of the Brillouin-Raman micro-spectroscopy (BRaMS) setup, consisting of a confocal microscope (CM-1), a Tandem Fabry-Perot interferometer (TFP-2 HC) and a Raman monochromator (RM) (see *Methods* for more details) (b) Scheme of the optical isolators used in the TFP-2 HC (side and front view) also represented by the small diode-like symbols in (a). The instrument uses circular polarizing techniques to prevent back reflections and so completely decouple the separate passes c) Spectral response of the interferometer. This yields an instrumental contrast better than 150 dB.

The tandem Fabry-Perot interferometer schematized in Fig.1, here presented for the first time (see *Methods*), reaches the unprecedented >150 dB contrast in just 3+3 pass configuration, thanks to the use of optical isolators, but also preserves a high luminosity, enhanced by the use of an avalanche-photodiode detector. This configuration enables one to exploit all the scattered light, to improve by more than 50 dB the contrast with respect to traditional 3+3 pass interferometers, making it possible to measure truly opaque samples and, eventually, to reduce the acquisition time to values typical of Raman spectrometers for joint Brillouin-Raman rapid raster-scan mapping and time-sampling experiments.

To demonstrate the performances of the new instrument we report the Brillouin-Raman mapping of a microbial biofilm. Biofilms are ubiquitous structures [24] formed by microbial cells growing onto solid surfaces embedded in a polymer matrix, often made of eso-polysaccharides, produced by the cells themselves.[25] City of microbes [26], with a complex morphology, the importance of biofilms depends on their increased resistance to antibiotics, anti-fungal drugs and extreme conditions.[4,27,28] Despite being poorly understood, the mechanical characteristics of the biofilm are of primary interest to elucidate the mechanisms governing the stability and the dispersion of the cells involved in the biofilm [29,30].

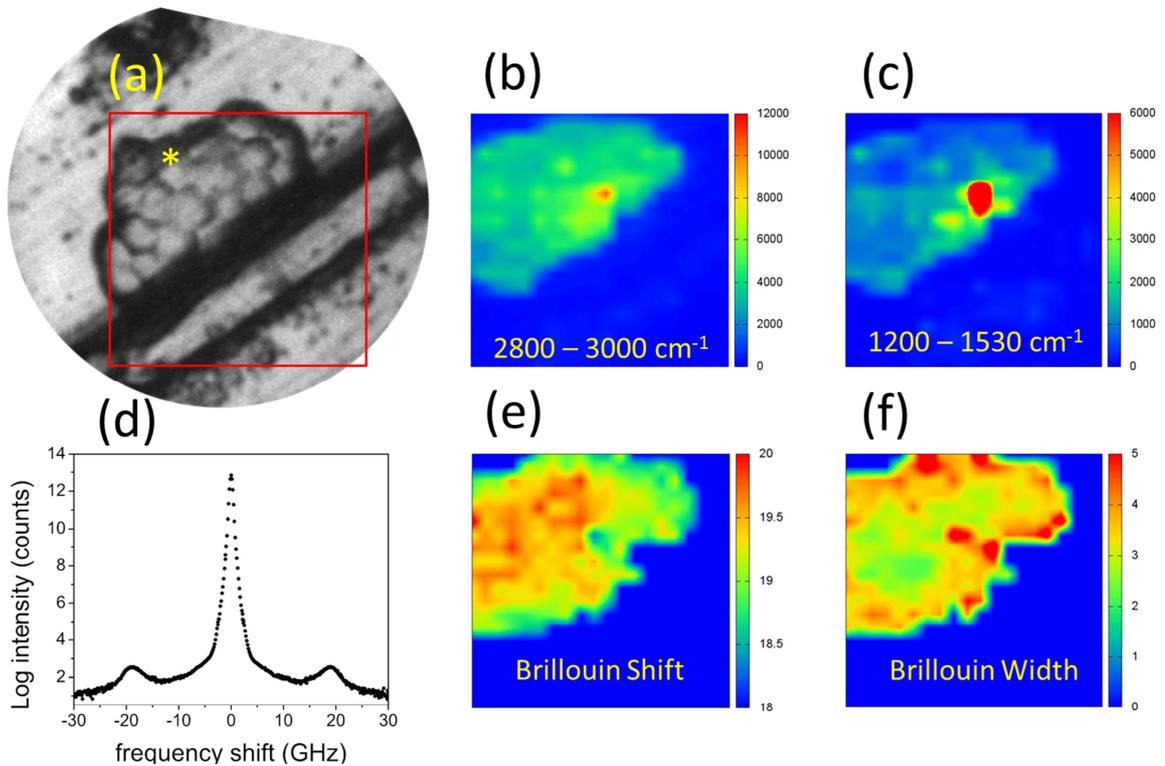

**Figure 2.** a) Optical micrograph of a *Candida albicans* biofilm. The red box denotes the 20μm × 20μm area whereby Brillouin-Raman raster-scan maps were obtained. b-c) Raman map based on the integrated intensity of the bands in the region indicated in the label. d) Brillouin spectrum extracted from the point denoted by an asterisk in a), showing that a minimum contrast of approximately 110dB is required to detect longitudinal phonons in this sample. The quasielastic signal (<9GHz) was obtained with the laser beam attenuated by calibrated neutral filters e-f) Brillouin map based on the characteristic frequency $\nu$ [GHz] and linewidth $\Gamma$ [GHz] of the Brillouin peak.

Fig.2a shows a photomicrograph of a *Candida albicans* biofilm in which a multilayer cluster of round shaped cells is visible. The Brillouin and Raman maps of the sample were obtained using a 50✖ objective, 1μm step, 20×20 points. Spectra were collected simultaneously, with an acquisition time of 10 s and an incident laser power of 17mW. Fig.2d shows that a contrast higher than 110dB is required to detect longitudinal phonons in this highly scattering sample. Raman maps were obtained by plotting the intensity of the bands in the range (Fig.5b) 2800-3000 $cm^{-1}$, showing the prominent $CH_2$ stretching mode of proteins and lipids, and (Fig.5c) 1200-1530 $cm^{-1}$, the so-called fingerprint region, comprising bands due to the amide III, cytochrome c, carbohydrates, proteins and lipids. The observed heterogeneity in concentration of CH groups across the scanned area can be attributed to a non-uniform thickness of the biofilm, with an increase in intensity by a factor 2-3 in the region of transition from one to two or more *Candida* layers. The most interesting feature is the high intensity of Raman bands in the central part of the biofilm. In this region, a prominent maximum in the bands at 1200-1530 $cm^{-1}$, whose intensity increases by a factor 20, cannot only be attributed to the multilayer structure of the biofilm, but also to a modification in the cell status and/or in the biofilm composition. In particular, we can expect the presence of zones where more extracellular polymeric substance is present and where the survival of Candida cells is favoured. In this respect, it is interesting to note that the out-of-scale increase of Raman signals in the range 1200-1530 $cm^{-1}$ could be at least partially attributed to the resonant scattering of cytocrome c, a marker of cell vitality.[31,32] Crucial to understand this feature is the noticeable change in the viscoelastic properties of the biofilm, visible in the Brillouin maps (Fig.2e-f). In the same central region, a ~6% reduction of the Brillouin frequency corresponding to a ~12% reduction in stiffness of the sample is observed, together with a twofold increase in linewidth (related to the acoustic attenuation in the sample), which is typical of heterogeneous viscoelastic media. The combination of reduced stiffness and increased acoustic attenuation has already been observed in dried biological tissues and attributed to the plasticizing effect associated with an increase of residual water.[15] In fact, both spectral changes observed here can indeed be attributed to a reduction of the structural relaxation time, i.e. of the local viscosity, associated to a local increase in the hydration level. Note that, in the absence of independent information on the local viscosity of the sample derived from the linewidth of Brillouin peaks, it is not possible to distinguish between the true, observed, relaxational effect and an alternative, fallacious, interpretation in terms of a merely elastic effect related to e.g. a reduction in microfilm connectivity. At the same time, without foundational molecular insight provided by simultaneous Raman spectra it is not possible to assign the observed micro-mechanical properties, which makes the fully integrated Brillouin-Raman setup needed and necessary.

The observed physicochemical heterogeneity (sample thickness, viscoelastic behavior and chemical modification) is compatible with the presence of a region with live *Candida* cells. This result is in line with the microbiological evidence that the biofilm acts as a structure that increases the resistance of cells to challenging agents. In fact, it is plausible that, in the region of larger thickness, the buried cells are protected by the overlaying biofilm structure. The protecting layer is also able to preserve water inside the inner cells, as suggested by the softer behavior revealed by Brillouin scattering. Our proof-of-concept investigation shows how an unprecedented high *contrast* together with a high spectral *resolution* and the *simultaneous* Brillouin-Raman mapping capability are able to merge in our instrument, and together open up new important investigation routes in microbiology, such as the in-situ detection and characterization of pernicious "persister cells", i.e those cells which are able to survive in the presence of devitalizing agents[33], and are amongst the major determinants of hospital infections.

A second experiment shows the potential of BRaMS measurements to monitor *fast* mechanical and chemical changes occurring in reacting samples. The production process of technologically important materials is often multi-stage [34] or involves complex thermal cycles of reaction. Here one major need for constructing mechanically optimized products is the synchronized determination of mechanical properties and extent of reaction, throughout the process. To test the capability of our system, Fig.3 shows the time evolution of Brillouin and Raman spectra during the isothermal polymerization of an epoxy-based thermoset resin, used as a matrix in high-performance composites [35].

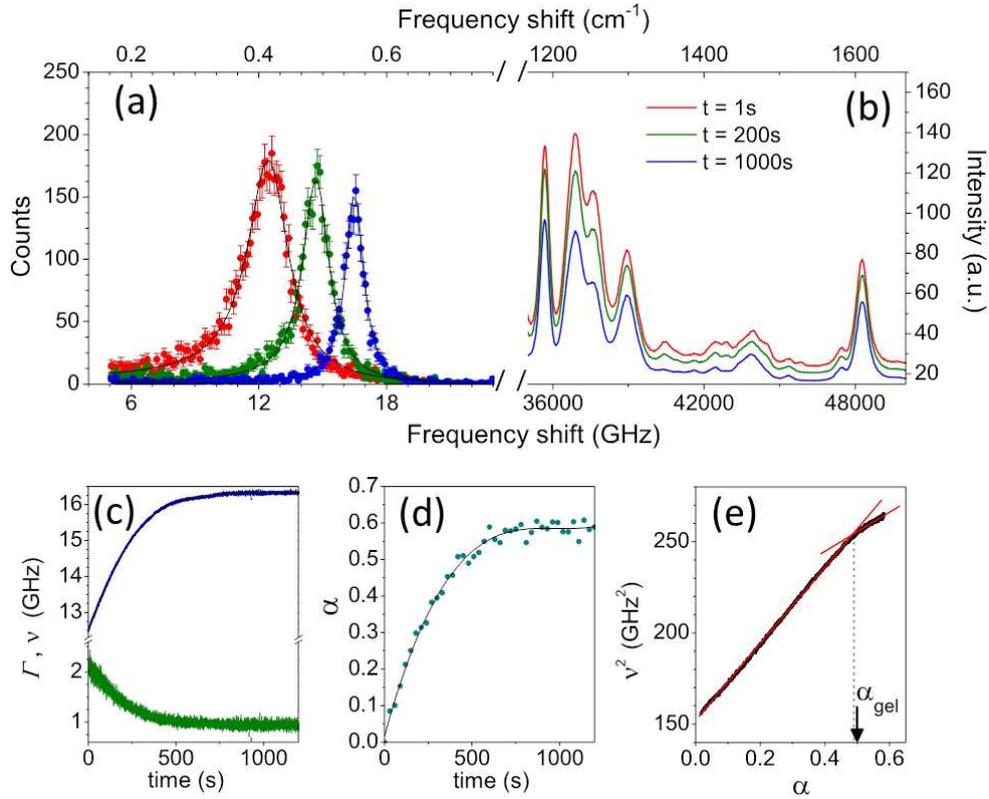

**Figure 3** (a) Brillouin and (b) Raman spectra of an epoxy-amine mixture (DGEBA-DETA 5:2) measured at different times of the isothermal polymerization process at T=65 °C: red symbols, 1 s; green symbols, 200 s; blue symbol, 1000 s since the beginning. Lines are the results of curve-fit analysis applied to the Brillouin peaks (see Methods). (c) Temporal evolution of the characteristic frequency ($\nu$) and linewidth ($\Gamma$) of Brillouin peaks. (d) Time evolution of the conversion $\alpha$ calculated from the intensity of Raman peaks (see text). (e) Plot of the square of the Brillouin frequency vs. $\alpha$. The sol-gel transition is expected to occur at $\alpha_{gel}$=0.5.

Brillouin spectra (Fig.3a) were taken every 0.5 s, with 30 mW laser intensity. The hardening of the resin is indicated by a progressive increase of the frequency shift $\nu$ and a parallel decrease of linewidth $\Gamma$ of the Brillouin peaks (Fig.3c); these parameters are related to the longitudinal elastic modulus and the longitudinal viscosity, respectively (see Methods). Indeed, the hardening of the resin results from the combination of two major effects, an increase of elastic energy density due to the formation of chemical bonds between DGEBA and DETA molecules and an increase in structural relaxation time of the system as it approaches the glass transition.

The molecular parameter used to monitor the advancement of the reaction is the conversion $\alpha$, i.e. the fraction of reacted epoxy rings during the curing process. Raman spectroscopy can directly access this parameter through the normalized intensity $I(t) = I_{1255}(t)/I_{1610}(t)$, where $I_{1255}(t)$ is the intensity of the epoxy band and $I_{1610}(t)$ is the intensity of the phenyl ring stretching band.[36] In fact, the epoxy groups are consumed during the reaction, while phenyl rings are unaffected by the curing process. The time evolution of the conversion, measured as $\alpha(t) = 1 - I(t)/I(0)$, is reported in Fig.3d together with a polynomial fit. A transition occurs at 8 minutes from the beginning, i.e. at about $\alpha = 0.49$, clearly visible in the evolution of $\nu^2$ vs. $\alpha$ (Fig.3). This phenomenon can be attributed to the sol-gel transition that, in the ideal case, is expected to occur at $\alpha_{gel} = 0.5$, according to the Flory-Stockmayer theory.[37,38] In the absence of relaxation processes, $\nu^2$ is proportional to the elastic modulus (see Methods) that, in turn, is proportional to the energy density of chemical bonds, i.e. to the number of bonds formed during the reaction. This condition is better satisfied in the second part of the reaction, and can explain the linear behavior of $\nu^2$ vs. $\alpha$ for $\alpha > 0.49$. Conversely, the presence of the structural relaxation and its evolution during the reaction[39], attested by the variation of $\Gamma$ in Fig.3c, is possibly responsible for the faster variation of $\nu^2$ with respect to $\alpha$ in the first part of the reaction. How the complex interplay between structural relaxation and bond formation can generate the linear behavior of $\nu^2$ vs. $\alpha$ for $\alpha < 0.49$ is still unclear. If this will be found to be general, the frequency of Brillouin peaks could be used as an effective probe for the advancement of a polymerization reaction.

To our knowledge, this is the first simultaneous measurement of chemical and mechanical evolution during a polymerization reaction. The proposed BRaMS setup is suitable for *in-situ* measurements, with fraction-of-second time resolution, in fraction-of-picoliter sample volumes, including those composite materials which, due to an exceedingly high elastic scattering, are out of reach of existing devices.

A third experiment benefits from the wide, tunable *spectral range* accessible to the tandem Fabry–Perot interferometer and from its polarization selectivity to perform extended depolarized light scattering (EDLS) investigations of biological matter.

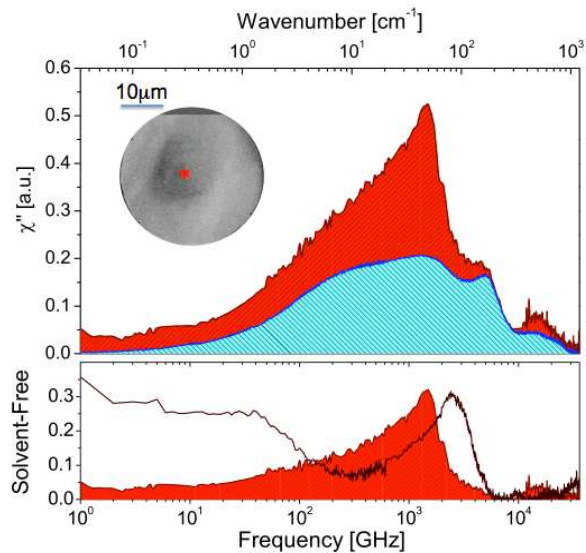

**Figure 4** Top: EDLS susceptibility of fixed NIH 3T3 fibroblast cell line (red profile) and PBS solution (cyan profile), at T = 25 °C. The broad-band spectral profiles have been obtained by the composition of five interferometric and one Raman spectrum. Inset: Image of NIH 3T3 single cell acquired with 50× Mitutoyo objective. Bottom: Solvent free spectrum obtained by subtracting the spectral profile of the PBS solvent from that of the cell (red profile). The solvent free spectrum of a 100 mg/ml lysozyme aqueous solution is also shown for comparison (brown line)[37].

EDLS has been recently developed as an implementation of depolarized Raman scattering to access both vibrational and relaxation dynamics of matter in the wide frequency range between fractions of GHz and tens of THz [12]. In this range, we can address a variety of hot topics of condensed matter, going from hydration of biological systems[40] to anharmonicity and boson peak in glasses[41] and highly viscous media.[13,42] EDLS measurements, till now performed by means of separate dispersive and interferometric set-ups, can be now obtained with a unique instrument, simultaneously and with high spatial resolution, paving the way to the development of micro-EDLS.

Fig.4 shows the low frequency depolarized Raman spectrum of a NIH/3T3 murine fibroblast, measured in the central region of the nucleus. The solvent-free (SF) spectrum, obtained after the subtraction of the solvent contribution, is also reported, together with the SF spectrum previously obtained by a conventional setup for the aqueous solution of a protein.[37] The spectrum from the nucleus of the cell is dominated by a strong vibrational contribution around 1THz. In addition to possible local vibrations in the complex structure of the nucleus, this contribution can be associated to the boson peak,[43] an ubiquitous signature of collective vibrations in disordered condensed matter[44]. Conversely, the strong contribution at 2-3THz in the spectrum of the protein is mainly attributed to the libration of methyl groups[45]. The number of methyl groups in DNA is much smaller than in proteins[46], and this can explain the suppression of the 2THz contribution in the spectrum of the nucleus. Worthy of note is also the GHz range, where the spectrum is noticeably less intense than in the water-protein solution. In fact, the relaxation processes in DNA are known to be about one order of magnitude slower than in proteins [46], hence their spectral signature is out of our spectral range. The residual intensity in the 10-100 GHz region can be attributed to the relaxation of hydration water, which is dynamically retarded by a factor greater than three with respect to bulk water.

To our knowledge, BRaMS gives the first evidence of boson peak and of relaxational contributions from hydration water within the nucleus of a single cell. These picosecond motions have been proposed to be responsible for directing biochemical reactions and energy transport in biological matter[47,48] and even to control the drug intercalation in DNA[49]. BRaMS opens the route to the study of these phenomena *in-situ* and, eventually, *in-vivo*.

In conclusion we have demonstrated *rapid* and *simultaneous* Brillouin and Raman experiments, with *high contrast*, *high resolution* and *extended spectral range*. The performances of the novel technique have been tested in three case studies: *i*) the *in situ* raster-scan mapping of viscoelastic and chemical properties of biofilms, with the potential of revealing persister cells; *ii*) the simultaneous monitoring of mechanical and chemical changes in reacting materials with sub-second time resolution; *iii*) the collective dynamics of biological matter by extended depolarized light scattering with subcellular spatial resolution. This noninvasive method is now available for a variety of applications, having potential for in vivo diagnosis of pathologies involving viscoelasticity changes as well as altered structure and composition.

## METHODS

**Experimental Setup**

Figure 1(a) shows a schematic of the experimental arrangement for Brillouin-Raman measurements. A 532nm single mode solid-state laser (SpectraPhysics Excelsior) was employed as light source (S). The vertically polarized laser beam is first sent through a JRS Scientific Instruments TCF-1 temperature controlled etalon, which reduces the intensity of laser side lobes by a factor of about 600, and then to the input of a customized JRS Scientific Instruments CM-1 Confocal Microscope, specifically developed in collaboration with the manufacturer to allow Raman/Brillouin signal splitting.

Inside the microscope, the laser beam is first expanded and collimated (BE), then a broadband polarizing beam splitter (BS) sends it to an infinity-corrected apochromatic microscopic objective. In the present experiments we used a Mitutoyo M-Plan Apo 20× with a very long working distance of 20 mm, a 2 μm depth of field and a limiting numerical aperture of 0.42, and a Mitutoyo M-Plan Apo 50× with a working distance of 13 mm and a numerical aperture of 0.55. The objective is used both to focus the laser beam onto the sample and to collect the back-scattered light. The laser spot diameter on the surface is ~2 μm measured with the 20× objective. Illumination of the sample surface is obtained by means of a coaxial LED illuminator, providing blue light peaked at about 470nm. A broadband λ/4 retarder wave plate can be inserted upstream of the objective to switch from depolarized to unpolarized scattering configuration.

The polarizing beamsplitter forwards the depolarized backscattered light towards the spectrometers. In the collimated beam section of the microscope, a tunable ultrasteep short-pass filter (TEF, Semrock SP01-561RU) transmits the anti-Stokes quasielastic scattered light to a JRS Scientific Instruments TFP-2 HC Fabry–Perot interferometer and reflects the Stokes deeply inelastic scattered light towards the Horiba iHR320 Triax Raman Spectrometer (RM). The Fabry–Perot Spectrometer includes a narrow band pass filter (NBF) to rule out broadband light signals, while a further

RazorEdge longpass Filter (LPF) improves the rejection of residual elastic contributions on the Raman signal line.

Given the different cross section for Raman and Brillouin phenomena, Brillouin spectrometers are usually equipped with the most sensitive detectors, namely photomultipliers operating in single photon regime. In recent times, technologic improvements led to the development of single photon avalanche photodiodes (SPAD), reaching much higher quantum efficiencies than conventional phototubes. In our setup, a LaserComponents COUNT®-10 SPAD detector is used as a sensor inside the TFP-2 HC spectrometer. This device provides a quantum efficiency of ~70% at 532 nm, together with a maximum noise rate of 10 cps. The achievable signal/noise ratio is comparable with that of classical phototube detectors, with an approximate five-fold gain in terms of data gathering speed.

Pin-holes at the entrance of the TFP spectrometer provide a confocal arrangement for microscopic imaging: the broadband light gathered by the objective is imaged onto a USB CMOS sensor camera (SC), which allows us to visualize the sample surface. The field of view diameter depends on the objective focal length: using the 20× objective, it is ~87 μm. The use of both a polarizer beamsplitter and a laser line OD6 notch filter before the camera enable the image to be seen with no loss of signal during the joint measurements.

For mapping measurements, the sample is mounted onto a PI 611-3S Nanocube XYZ closed loop piezoelectric translation stage (SH), with resolution of 1 nm and a motion range of 100 μm per axis. The high voltage needed by the positioner is generated by a PI E-664 controller, voltage controlled through a homemade DAC/802.11 wireless control board. A software scripting system was developed starting from the JRS Automated Measurement application (JRS-AM) in order to execute sequences of simultaneous Raman/Brillouin measurements and stage positioning operations. LabSpec 5 software was used for acquisition of the Raman data and JRS GHOST 7 for the Brillouin data.

The TFP-2 HC is a very high contrast variant of the original Sandercock type tandem multi-pass interferometer[50]. The tandem configuration considerably increases the spectral range available for Brillouin and depolarized light scattering experiments, through the suppression of the replica of inelastic signals due to the side orders. In the high contrast model, polarization and retardation optics (Fig.1b) are used at each interferometer pass to effectively decouple each pass from the next one, improving the extinction by a factor higher than $10^5$. Before each passage through a Fabry-Perot, light passes through a 45° polarizer and is back reflected with circular polarization by a triangular prism. Light reflected from the next optical surface has opposite chirality and emerges at the polarizer with crossed polarization and is stopped. The adoption of this method also allows the manufacturer to adopt a perfectly orthogonal incidence of light on all surfaces, thus aiming at the theoretical contrast limit of $10^{18}$.

The laser spectrum in Fig. 1c was measured normalizing and merging several spectra of a laser beam at different values of the input power. During these measurements, the laser spectrum side lobes were attenuated by means of a series of two JRS TCF-1 temperature controlled etalons, having a FSR of about 68 GHz, both tuned to the laser frequency. The mirror spacing of the spectrometer was set at 2.20 mm in order to match the FSR of the etalon and small input and output pinholes were used, to reduce contributions from uncollimated light. The incoming ~70 mW laser beam was attenuated by means of a set of neutral filters and a variable one.

Further investigation are in progress to understand the origin of the residual features visible in Fig. 1c at the $10^{-15}$ level. The tiny inelastic structures may be attributed to residual side lobes of the original laser spectrum and, eventually, to spurious Brillouin scattering from optical components in the first part of the spectrometer. We can thus anticipate that the true contrast of the interferometer is greater than 150dB.

Moreover, it has to be noticed that the contrast (Fig.1) was measured for vertically polarized light, because of the selectivity of the instrument in polarization. Supposing to change the polarization input, as for the effect of depolarized elastic scattering, the 5 polarizers of internal groups of TFP-2 would strongly suppress the light intensity, while the instrumental contrast (related only to the Fabry-Perot mirrors reflectance) would remain the same. This selectivity makes the instrument even more suited for the study of biological samples, because naturally less sensitive to depolarized elastic scattering.

We can summarize the characteristics of the new setup as follows:

*Contrast and luminosity*: The 150 dB contrast documented in Fig. 1c was made possible not only by the six pass geometry of the interferometer, but also by the adoption of multiple optical isolators inside the device (Fig.1b) which prevent back propagation of light inside the interferometer. The unprecedented contrast reached by the TFP-2 HC is by far superior to that of previous 3+3 pass tandem Fabry-Perots (~100 dB for the TFP1).[50] In addition, the luminosity of our setup is ~0.2 which, together with the factor five improvement in the efficiency of the detector, opens up the possibility to map the viscoelastic properties of truly opaque or even reflecting samples.

*Resolution*: The full-width-at-half-height of the elastic line (Fig.1c) gives the spectral resolution, which is related to the FPs' mirror spacing and to their optical quality. With mirror spacing of 15 mm or higher, a value of linewidth better than 0.1 GHz can be obtained. This is more than a factor 7 better than in VIPAs, enabling the study of acoustic attenuation and relaxation processes in viscoelastic media.

*Acquisition time*: The high efficiency of the setup in collecting quasielastic scattered light, by using a polarizing cube and an edge filter as beam-splitters and an avalanche photodiode as photodetector, enables the acquisition time for a single spectrum to be reduced to seconds. It is thus possible to measure within a reasonable time both collective (by BLS) and single particle (by RS) properties of materials that are heterogeneous in space (e.g. biological matter or geological samples) or changing over time (e.g. reactive or degradable systems).

*Spectral range*: The tandem configuration is very effective in suppressing (better than 40 dB) the replicas of the transmission peaks of the two FP interferometers. By combining spectra recorded with mirror spacing in the range of 20 to 0.2 mm, a spectral range from fractions of GHz to more than 1 THz can be obtained. The spectral range is further extended to the high frequency side by means of the Raman spectrometer, which can operate in the range 0.9 to 100 THz. The possibility of collecting a continuous spectrum in such a wide and informative range is not possible with other setups, even with the most recent Raman-Brillouin ones[22] due to the very limited spectral range accessible to VIPAs.

**Analysis of Brillouin Spectra**

Brillouin spectra presented in Figs. 2 and 3 originate from light inelastically scattered by thermally activated acoustic modes. In a back scattering experiment from ideal homogeneous and elastic solids, two single Brillouin peaks can be revealed at frequency shifts $\nu$ given by $\omega = 2\pi\nu = \pm V \cdot q$, where $V$ is the velocity of the longitudinal acoustic modes, $q = 2nk_i$ is the momentum exchanged in the scattering process, $n$ is the refractive index of the sample and $k_i$ the wavevector of incident light. In ideally elastic samples, the longitudinal elastic modulus $M$ is simply given by $M = \rho V^2$, where $\rho$ is the mass density of the sample. In case of viscoelastic materials, damping mechanisms reduce the lifetime of acoustic modes and the Brillouin spectrum can be described by a damped harmonic oscillator (DHO) function:

$$I(\omega) = \frac{I_0}{\pi} \cdot \frac{\omega_b^2 \Gamma_b}{(\omega^2 - \omega_b^2)^2 + \omega^2 \Gamma_b^2} \tag{1}$$

where $\omega_b$ and $\Gamma_b$ approximately give the frequency position and the linewidth (FWHM) of Brillouin peaks. These parameters are related to the real ($M'$) and imaginary ($M'' = \omega\eta_L$) parts of the longitudinal elastic modulus through the relations:

$$M' = \rho\omega_b^2/q^2 \quad ; \quad \eta_L = \rho\Gamma_b/q^2 \tag{2}$$

The simple hydrodynamic model described by Eq.1, convoluted with the instrumental function, is appropriate to fit the whole Brillouin spectrum of materials (liquids or solids) having internal relaxation rates much faster than $\omega_b$. In case of relaxations close to $\omega_b$, generalized hydrodynamic models can be used to model the spectrum, introducing a frequency dependence either of the modulus $M$ or, equivalently, of the longitudinal viscosity $\eta_L$.[51] Luckily enough, also in this case Eq.1 can be used to fit the spectrum in correspondence to the peak maximum and, through Eq.2, obtain the real and imaginary part of the modulus at the single frequency of the Brillouin peak.[52] This fast procedure is the most appropriate for mapping purposes. Conversely, long and accurate measurements would be needed to obtain the shape of the relaxation functions from the band shape of Brillouin spectra, extending far from the resonance.[52]

**Analysis of EDLS Spectra**

The total EDLS spectrum reported in Fig.4 has been obtained by joining the experimental profiles collected by means of the interferometric and dispersive setups.[12] The interferometric portion of the spectrum (1-2000 GHz) derives from the combination of different mirror separations, namely $d$=14, 4, 1, 0.5 and 0.2 mm; for this last distance two free spectral range have been used in order to reach the thousands of GHz, and the overlapping with the dispersive segment. The high frequency part of the spectrum (from about 900 GHz) has been recorded by using a 1800 *gr/mm* grating with an aperture slit of 300 μm at the entrance of the monochromator. After the removal of the dark count contributions, low and high frequency portions have been spliced, benefiting from a large overlapping. Further, the imaginary part of the dynamic susceptibility $\chi''(\nu)$ has been derived as the ratio between the intensity of depolarized light $I_{VH}(\nu)$ and $[n(\nu) + 1]$ where $n(\nu)$ is the Bose – Einstein occupation number, namely $n(\nu) = \exp{[(h\nu/k_BT) - 1]^{-1}}$.

**Materials**

*Candida albicans* CMC 1768 strain used in this study was isolated from abdominal fluid of a patient in Intensive Care Unit in Pisa Hospital. The identification was carried out by comparing the ITS and LSU D1/D2 sequence with GenBank (http://blast.ncbi.nlm.nih.gov/Blast.cgi) and specific databases.[53,54] Biofilms were obtained by submerging disks of aluminum foil (4mm diameter) in Petri dishes containing YEPD (Yeast Extract 1%, Peptone 1%, Dextrose 2% - Difco Laboratories, Detroit, MI, USA) medium inoculated with yeast cells at $OD_{600}$=0.2. Petri dishes were placed in a rocking incubator at 37° C for 72 h, and then samples were washed twice with distilled sterile water and dried at room temperature for one week prior to analysis.

The epoxy-amine mixture used in this study is made of diglycidyl ether of bisphenol A (DGEBA, molecular weight 348) and diethylenetriamine (DETA, purity ≥99%) in the 5:2 stoichiometric ratio. The two monomer types are mutually reactive and polymerize by stepwise addition of the amino hydrogen to the epoxy group, with a rate of reaction strongly controlled by the temperature. The mixture was prepared by mixing for 2 min at room temperature and then transferred into the measurement cell, kept at a fixed T=65°C. Because of the presence of a multifunctional reagent, the reaction of DGEBA with DETA grows branched molecules and finally yields a network-polymer glass. Before reaching the vitrification point, the system goes through the gel point, where a percolating network of bonded particles has developed. The extent of reaction at the gel point, according to the classical theory of Flory-Stockmayer is $\alpha = 0.50$ for our system.

NIH/3T3 murine fibroblast cell line was purchased from American Type Culture Collection (ATCC). Cells were grown in Dulbecco Modified Eagle's Medium (DMEM) containing 10% (v/v) heat-inactivated fetal bovine serum (FBS), 100 U/mL penicillin, 100 U/mL streptomycin (SIGMA Aldrich, St. Louis) and maintained at 37 °C in a 5% CO humidified atmosphere. Cells were seeded in onto $CaF_2$ substrates sterilized using 100% ethanol and UV irradiation, treated with Poly-L-lysine solution. Fixation was performed by incubating cells with 4% paraformaldehyde in phosphate buffer saline (PBS) solution for 10 minutes at room temperature, then cells were washed twice with PBS.

**ACKNOWLEDGMENTS**

S.Co. acknowledges financial support from MIUR-PRIN (Project 2012J8X57P). S.Ca. acknowledges the support from PAT (Autonomous Province of Trento) (GP/PAT/2012) "Grandi Progetti 2012" Project "MaDEleNA". The authors acknowledge Jacopo Scarponi for valuable help in setting up the hardware and software system for simultaneous Raman and BLS measurements.

Karunarathna, U. Kõljalg, G. M. Kovács, E. Kraichak, K. Krizsan, C. P. Kurtzman, K.-H. Larsson, S. Leavitt, P. M. Letcher, K. Liimatainen, J.-K. Liu, D. J. Lodge, J. Jennifer Luangsa-ard, H. T. Lumbsch, S. S. Maharachchikumbura, D. Manamgoda, M. P. Martín, A. M. Minnis, J.-M. Moncalvo, G. Mulè, K. K. Nakasone, T. Niskanen, I. Olariaga, T. Papp, T. Petkovits, R. Pino-Bodas, M. J. Powell, H. A. Raja, D. Redecker, J. M. Sarmiento-Ramirez, K. A. Seifert, B. Shrestha, S. Stenroos, B. Stielow, S.-O. Suh, K. Tanaka, L. Tedersoo, M. T. Telleria, D. Udayanga, W. A. Untereiner, J. Diéguez Uribeondo, K. V. Subbarao, C. Vágvölgyi, C. Visagie, K. Voigt, D. M. Walker, B. S. Weir, M. Weiß, N. N. Wijayawardene, M. J. Wingfield, J. P. Xu, Z. L. Yang, N. Zhang, W.-Y. Zhuang, and S. Federhen, Database: the Journal of Biological Databases and Curation **2014**, bau061 (2014).

18